\documentclass[aps,pre,twocolumn,eqsecnum,showpacs,superscriptaddress]{revtex4}
\usepackage{graphicx,color}
\usepackage{amsmath}
\usepackage{amssymb}
\usepackage{bm}

\definecolor{dgreen}{rgb}{0,0.5,0}

\definecolor{delete}{cmyk}{0.5,0,0,0}

\begin{document}

\title{Quasi adiabatic dynamics of energy eigenstates for solvable quantum system at finite temperature}



\author{Takaaki Monnai}
\affiliation{Department of Materials and Life Sciences, Seikei University, Tokyo 180-8633, Japan}



\date{\today}

\begin{abstract}
It is a fundamental problem to characterize the nonequilibrium processes. For a slowly moving one-dimensional potential, we explore the quasi adiabatic dynamics of the initial energy eigenstates for a confined quantum system interacting with a large reservoir. For concreteness, we investigate a dragged harmonic oscillator linearly interacting with an assembly of harmonic oscillators, and explore the deviation from adiabatic processes by rigorously calculating the so-called persistent amplitude. In this way, we also show that the phase of the persistent amplitudes are common both for the ground and excited states. 
\end{abstract}       


\maketitle
\section{Introduction}
Recently, considerable attention has been paid to thermodynamic aspects of the nonequilibrium processes of many-body systems. Remarkable progresses include some universal relations such as the fluctuation theorems\cite{Evans1,Gallavotti1,Esposito1,Utsumi1,Andrieux1,Tameem1,Monnai1}, an energetics of mesoscopic systems\cite{Sekimoto1,Wang1,Seifert1,Roncaglia1,Nakamura1}, relaxation of thermally isolated quantum systems\cite{Hazibabich1,Kinoshita1,Rigol1}, to name but a few.   

By calculating the fidelity or so-called persistent amplitude, we can evaluate the relaxation time of isolated quantum many-body systems\cite{Monnai3,Santos1,Reimann1}. 
On the other hand, we can also calculate the time evolution of each energy eigenstate itself for adiabatic processes. 
In this article, we explore how the time evolution of initial eigenstates deviates from those of adiabatic processes for an externally perturbed quantum system interacting with a large reservoir by calculating the persistent amplitude.

For concreteness, we consider a uniformly dragged harmonic potential interacting with a reservoir which is an assembly of infinitely many harmonic oscillators. Such a model is useful to discuss a quantum system coupled to an environment: quantum Langevin equation\cite{Ford1,Zwanzig1}, atoms interacting with an electric field\cite{Gardiner1}, exact case studies of the quantum fluctuation theorem\cite{Monnai1,Monnai2}, and so on.

This paper is organized as follows.
In Sec. II, we describe our model.
In Sec. III, we calculate the persistent amplitude of the ground state in terms of the Wick's theorem.
In Sec. IV, we explore the excited states by considering the case of finite temperature.
Sec. V is devoted to a summary.
\section{Model}
In this section, we describe our model, and diagonalize the Hamiltonian.
We consider a harmonic potential linearly interacting with an assembly of harmonic oscillators\cite{Zwanzig1,Monnai1,Gardiner1,Monnai2}. We externally control the center of the potential $f(t)$.  
Then, the total Hamiltonian is 
\begin{equation}
\hat{H}(t)=\frac{\hat{p}^2}{2m}+\frac{k}{2}(\hat{q}-f(t))^2+\frac{1}{2}\int d\lambda(\hat{p}_\lambda^2+\omega_\lambda^2 (\hat{q}_\lambda-\kappa_\lambda \hat{q})^2). \label{Hamiltonian1}
\end{equation}
Here, $m$ is the mass and $k$ stands for the spring constant. Also, $\hat{q}$ and $\hat{p}$ are the position and momentum of a particle, and $\hat{q}_\lambda$, and $\hat{p}_\lambda$ are those of the reservoir degrees of freedom. 
These operators satisfy the canonical commutation relations $[\hat{q},\hat{p}]=i\hbar$, $[\hat{q}_\lambda,\hat{p}_{\lambda'}]=i\hbar\delta(\lambda-\lambda')$, and the system variables commute with those of the reservoir. We assume that the coupling strength $\kappa_\lambda$ between the system and the reservoir is weak so that the Hamiltonian does not admit any bound states.  
It is convenient to define the normal mode for the reservoir, $\hat{a}_\lambda=\frac{1}{\sqrt{2\hbar\omega_\lambda}}(\omega_\lambda \hat{q}_\lambda+i\hat{p}_\lambda)$.
Then, the total Hamiltonian is diagonalized as
\begin{eqnarray}
&&\hat{H}(t) \nonumber \\
&=&\hat{H}_0+\hat{V}_I(t) \nonumber \\
&=&\int d\lambda\hbar\omega_\lambda\hat{A}_\lambda^\dagger\hat{A}_\lambda \nonumber \\
&&+\sqrt{\frac{\hbar}{2}}kf(t)\int d\lambda(\frac{\kappa_\lambda\omega_\lambda\sqrt{\omega_\lambda}}{\eta_-(\omega_\lambda)}\hat{A}_\lambda+\frac{ \kappa_\lambda\omega_\lambda\sqrt{\omega_\lambda}}{\eta_+(\omega_\lambda)}\hat{A}_\lambda^\dagger) \nonumber \\
&&+\frac{k}{2}f(t)^2. \label{Hamiltonian2} 
\end{eqnarray}
Here, the normal mode is a linear combination of canonical operators $\hat{A}_\lambda=\hat{a}_\lambda-\frac{\kappa_\lambda\omega_\lambda\sqrt{\omega_\lambda}}{\eta_+(\omega_\lambda)}\{\frac{m\omega_\lambda\hat{q}+i\hat{p}}{\sqrt{2\hbar}}+\int\frac{d\lambda'}{2}\frac{\kappa_{\lambda'}\omega_{\lambda'}\sqrt{\omega_{\lambda'}}\hat{a}_{\lambda'}}{\omega_\lambda-\omega_{\lambda'}+i0}+\frac{\kappa_{\lambda'}\omega_{\lambda'}\sqrt{\omega_{\lambda'}}\hat{a}_{\lambda'}^\dagger}{\omega_\lambda+\omega_{\lambda'}})\}$. 
Note that we omitted the vacuum energy, which is time-independent.
The normal mode is obtained by diagonalizing the Hamiltonian $\hat{H}$ in the absence of $f(t)$ in (\ref{Hamiltonian1}) as $\hat{H}=\int d\lambda\hbar\omega_\lambda\hat{A}_\lambda^\dagger\hat{A}_\lambda$.
The normal mode satisfies the canonical commutation relation $[\hat{A}_\lambda,\hat{A}_{\lambda'}^\dagger]=\delta(\lambda-\lambda')$.
Here, we introduced the dispersion function $\eta_\pm(z)=mz^2-k-\int d\lambda\kappa_\lambda^2\omega_\lambda^2-\int d\lambda\frac{\kappa_\lambda^2\omega_\lambda^4}{z^2-\omega_\lambda^2\pm i0}$.   
To calculate the persistent amplitude in the following sections, it is convenient to use the interaction picture.
The interaction Hamiltonian in the interaction picture is given as
\begin{eqnarray}
&&\hat{H}_I(t) \nonumber \\
&=&\sqrt{\frac{\hbar}{2}}kf(t)\int d\lambda(\frac{\kappa_\lambda\omega_\lambda\sqrt{\omega_\lambda}}{\eta_-(\omega_\lambda)}\hat{A}_\lambda e^{-i\omega_\lambda t} \nonumber \\
&&+\frac{\kappa_\lambda\omega_\lambda\sqrt{\omega_\lambda}}{\eta_+(\omega_\lambda)}\hat{A}_\lambda^\dagger e^{i\omega_\lambda t})+\frac{k}{2}f(t)^2. \label{Interaction1}
\end{eqnarray}
\section{Persistent amplitude} 
In this section, we recapitulate the way to calculate the vacuum persistent amplitude for the ground state $|0\rangle$. At $t=0$, we assume that the interaction and Heisenberg pictures coincide, and their vacuum states are common. 
Instead of the adiabatic switching-off of the interaction in the scattering theory, we move the center of the trap sufficiently slowly in a cyclic way $f(T)=f(0)$ for a long time $T$ as shown in Fig. 1. The sign of the velocity $\dot{f}(t)$ is changed at $t=\frac{T}{2}$, however, thermodynamic properties of the stationary state are considered to depend only on the absolute value $|\dot{f}(t)|$.      
\begin{figure}
\center{
\includegraphics[scale=0.6]{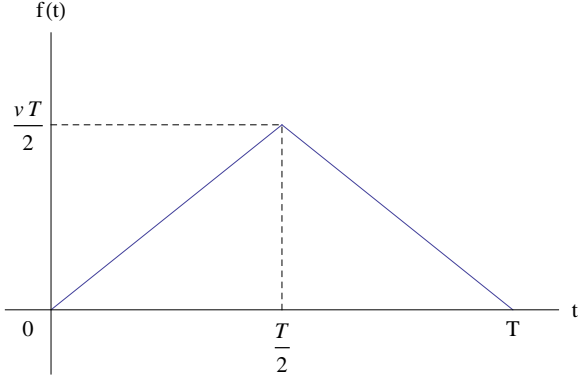}
}
\caption{(Color online)Cyclic manipulation of the center of the potential. The blue line shows the uniform dragging case $f(t)=v t$ for $0\leq t\leq\frac{T}{2}$ and $f(t)=v(T-\frac{t}{2})$ for $\frac{T}{2}\leq t\leq T$.}
\end{figure}  
According to the adiabatic theorem\cite{Thirring1}, the initial vacuum eigenstate $|0\rangle$ is considered to evolve toward a state $U(t)|0\rangle$ which is close to the eigenstate of $\hat{H}(t)$ for sufficiently slow processes. Note that the adiabatic theorem requires non-degenerated eigenstates, while our calculation similarly holds for the case of discrete spectrum by replacing the reservoir Hamiltonian with the discrete one.   

The time evolution operator in the interaction picture $U(t)={\rm T}\{e^{-\frac{i}{\hbar}\int_0^t ds \hat{H}_I(s)}\}$ is defined as a solution of $i\hbar\frac{\partial}{\partial t}U(t)=\hat{H}_I(t)U(t)$ under the initial condition $U(0)=1$. Here, ${\rm T}\{\cdot\}$ stands for the time-ordered product.
With the use of the evolution operator in the Heisenberg picture, $U(t)$ is rewritten as 
\begin{equation}
U(t)=e^{\frac{i}{\hbar}\hat{H}_0 t}{\rm T}\{e^{-\frac{i}{\hbar}\int_0^t ds \hat{H}(s)}\}.  \label{unitary1}
\end{equation}  

Let us calculate the vacuum persistent amplitude\cite{Zuber1}
\begin{equation}
\langle 0|{\rm T}\{e^{-\frac{i}{\hbar}\int_0^T dt \hat{H}_I(t)}\}|0\rangle, \label{amplitude1}
\end{equation}
which measures a distance between the initial and final states.     
For an adiabatic switching of $f(t)$, (\ref{amplitude1}) is equal to 
\begin{equation}
e^{-\frac{i}{\hbar}\Delta E T} \label{amplitude2}
\end{equation} 
with the use of (\ref{unitary1}).
Here, $\Delta E$ is the difference between the initial and final eigenenergies. 
Indeed, we have from (\ref{unitary1}) $\langle 0|e^{\frac{i}{\hbar}\hat{H}_0 T}{\rm T}\{e^{-\frac{i}{\hbar}\int_0^T dt \hat{H}(t)}\}|0\rangle=e^{\frac{i}{\hbar}E_0 T}\langle 0|{\rm T}\{e^{-\frac{i}{\hbar}\int_0^T dt \hat{H}(t)}\}|0\rangle$ with the eigenenergy $E_0$ of $\hat{H}(0)$, and the adiabatic evolution makes the initial state $|0\rangle$ to a state close enough to the corresponding ground state of $\hat{H}(t)$ for $0\ll t\leq T$.
For quasi adiabatic nonequilibrium processes, however, the final state is out of equilibrium and not an eigenstate. 
The absolute value of the persistent amplitude (\ref{amplitude1}) is close to unity for quasi adiabatic processes. 
We note that the eigenenergies of $\hat{H}(t)$ and $\hat{H}(0)$ are the same, since they are related by a unitary operator $\hat{H}(t)=D(t)\hat{H}(0)D(t)^\dagger$ with $D(t)=e^{-\frac{k}{\sqrt{2\hbar}}\int d\lambda(\frac{\kappa_\lambda\sqrt{\omega_\lambda}}{\eta_-(\omega_\lambda)}f(t)\hat{A}_\lambda-\frac{\kappa_\lambda\sqrt{\omega_\lambda}}{\eta_+(\omega_\lambda)}f(t)\hat{A}_\lambda^\dagger})$. This invariance is specific to our model, however, it is compatible with the adiabatic theorem for $\Delta E=0$ and $v=0$.  The phase shift is caused by a nonequilibrium deviation of $U(t)|0\rangle$ from the corresponding eigenstate.  

The vacuum persistent amplitude is further calculated as
\begin{equation}
\langle 0|{\rm T}\{e^{-\frac{i}{\hbar}\int_0^T dt\hat{V}(t)}\}|0\rangle e^{-\frac{i}{2\hbar}k\int dt f(t)^2}, \label{amplitude3}
\end{equation}
where we defined the interaction Hamiltonian minus the energy stored in the harmonic potential $\hat{V}(t)=\hat{H}_I(t)-\frac{k}{2}f(t)^2$.

We use the Wick's theorem\cite{Zuber1}
\begin{eqnarray}
&&{\rm T}\{e^{-\frac{i}{\hbar}\int_0^T \hat{V}(t)dt}\} \nonumber \\
&=&{\rm N}\{e^{-\frac{i}{\hbar}\int_0^T \hat{V}(t)dt}\}e^{-\frac{1}{2\hbar^2}\int_0^T dt_1\int_0^T dt_2\langle 0|{\rm T}\{\hat{V}(t_1)\hat{V}(t_2)\}|0\rangle}. \nonumber \\
&& \label{Wick1}
\end{eqnarray}
Then, the phase of the persistent amplitude $\Theta$ is related to the propagator 
\begin{eqnarray}
&&e^{i \Theta} \nonumber \\
&=&e^{-\frac{1}{2\hbar^2}\int_0^T dt_1\int_0^T dt_2\langle 0|{\rm T}\{\hat{V}(t_1)\hat{V}(t_2)\}|0\rangle}e^{-\frac{i}{\hbar}\frac{k}{2}\int dt f(t)^2},  \nonumber \\
&&\label{phase1}
\end{eqnarray}
since the vacuum expectation value of the normal ordered product is unity. 
In (\ref{phase1}), the propagator is calculated as
\begin{eqnarray}
&&\langle 0|{\rm T}\{\hat{V}(t_1)\hat{V}(t_2)\}|0\rangle \nonumber \\
&=&\frac{\hbar k^2}{2}f(t_1)f(t_2)\int d\lambda\frac{\kappa_\lambda^2\omega_\lambda^3}{|\eta_+(\omega_\lambda)|^2} e^{-i\omega_\lambda|t_1-t_2|}. \label{propagator1}
\end{eqnarray}
From Eqs.(\ref{phase1},\ref{propagator1}), the phase $\Theta$ is given as
\begin{eqnarray}
&&\Theta \nonumber \\
&=&(i\frac{k^2}{4\hbar}\int_0^T dt_1\int_0^T dt_2 f(t_1)f(t_2) \nonumber \\
&&\int d\lambda\frac{\kappa_\lambda^2\omega_\lambda^3}{|\eta_+(\omega_\lambda)|^2}e^{-i\omega|t_1-t_2|}-\frac{k}{2\hbar}\int_0^T dt f(t)^2)T+{\cal O}(1), \nonumber \\
&&\label{energyshift1}
\end{eqnarray}
where the ${\cal O}(1)$ contribution is negligible for the quasi adiabatic processes and calculation of the phase shift per unit time. 
Note that the first term of the right hand side of (\ref{energyshift1}) is actually real as shown for our model in the following section. 
\section{Uniform dragging}
In this section, we calculate the phase (\ref{energyshift1}) for the uniform dragging case, which is shown in Fig. 1. 
 
For the case of uniform dragging $f(t)=vt$ for $0\leq t\leq \frac{T}{2}$ and $f(t)=v(T-t)$ for $\frac{T}{2}\leq t\leq T$, the phase is then evaluated as
\begin{eqnarray}
&&\Theta \nonumber \\
&=&-\frac{k}{24\hbar}v^2 T^3+\frac{k^2v^2}{24\hbar}\int d\lambda\frac{\kappa_\lambda^2\omega_\lambda^2}{|\eta_+(\omega_\lambda)|^2}T^3 \nonumber \\
&&+\frac{k^2v^2}{2\hbar}\int d\lambda\frac{\kappa_\lambda^2}{|\eta_+(\omega_\lambda)|^2}T+{\cal O}(v^2). \label{energyshift2} 
\end{eqnarray}   
Remarkably, the absolute value of the persistent amplitude (\ref{amplitude1}) $e^{-\frac{4k^2 v^2}{\hbar}\int d\lambda\frac{\kappa_\lambda^2}{|\eta_+(\omega_\lambda)|^2\omega_\lambda}\sin^4\frac{\omega_\lambda T}{4}}$
 converges to unity in the quasi static limit $v\rightarrow 0$, which is consistent with the adiabatic theorem.  

With the use of the lemma $\int d\lambda\frac{\kappa_\lambda^2\omega_\lambda^2}{|\eta_+(\omega_\lambda)|^2}=\frac{1}{k}$ detailed in the appendix, the first and second terms of (\ref{energyshift2}) cancel each other, and we can further calculate the phase as
\begin{equation}
\Theta=\frac{k^2v^2}{2\hbar}\int d\lambda\frac{\kappa_\lambda^2}{|\eta_+(\omega_\lambda)|^2}T+{\cal O}(v^2). \label{energyshift3}
\end{equation}
Here, we have some remarks.   
First, the phase shift $\Delta \Theta$ is proportional to $T$, positive, 
and quadratic function of the velocity.
On the other hand, the absolute value of the persistent amplitude exponentially decays for fast perturbations.
In this case, the dragging is no longer corresponding to a quasi adiabatic process.

\section{Finite temperature}
In this section, we explore the case of excited states.
We show that the phase of the excited states are the same as that of the ground state. 
For this purpose, let us consider the initial canonical state $\hat{\rho}_c=\frac{1}{Z}e^{-\beta\hat{H}(0)}$ at an inverse temperature $\beta$. Here, $Z={\rm Tr}e^{-\beta\hat{H}(0)}$ is the partition function. 
We calculate the persistent amplitude for the canonical state $\langle {\rm T}\{e^{-\frac{i}{\hbar}\int_0^T\hat{H}_I(t)dt}\}\rangle_c={\rm Tr}\hat{\rho}_c{\rm T}\{e^{-i\frac{i}{\hbar}\int_0^T\hat{H}_I(t)dt}\}$.
With the use of the initial energy eigenstate $|E_\mu\rangle$, the persistent amplitude of $\hat{\rho}_c$ is equal to    
\begin{eqnarray}
&&\langle{\rm T}\{e^{-\frac{i}{\hbar}\int_0^T\hat{H}_I(t)dt}\}\rangle_c \nonumber \\
&=&\int d\mu \frac{1}{Z}e^{-\beta E_\mu}\langle E_\mu|{\rm T}\{e^{-\frac{i}{\hbar}\int_0^T\hat{H}_I(t)dt}\}|E_\mu\rangle, \label{energyshift5}
\end{eqnarray}
where $\mu$ labels the excited states.
In particular, we calculate the phase $\Theta_\mu$ for $\langle E_\mu|{\rm T}\{e^{-\frac{i}{\hbar}\int_0^T\hat{H}_I(t)dt}\}|E_\mu\rangle$. 
 
We note that the normal ordering can be decomposed as
\begin{equation}
{\rm N}\{e^{-\frac{i}{\hbar}\int_0^T\hat{V}(t)dt}\}=e^{-\frac{i}{\hbar}\int_0^T\hat{V}^{(-)}(t)dt}e^{-\frac{i}{\hbar}\int_0^T\hat{V}^{(+)}(t)dt}, \label{factor1}
\end{equation}
which is shown by expanding both sides.
Here, we introduced field operators $\hat{V}^{(+)}(t)=-\sqrt{\frac{\hbar}{2}}kf(t)\int d\lambda\frac{\kappa_\lambda\omega_\lambda\sqrt{\omega_\lambda}}{\eta_-(\omega_\lambda)}\hat{A}_\lambda e^{-i\omega_\lambda t}$
and $\hat{V}^{(-)}(t)=-\sqrt{\frac{\hbar}{2}}kf(t)\int d\lambda\frac{\kappa_\lambda\omega_\lambda\sqrt{\omega_\lambda}}{\eta_-(\omega_\lambda)}\hat{A}_\lambda^\dagger e^{i\omega_\lambda t}$.
We also use a lemma\cite{Monnai2,Thirring1,Banchi1}
\begin{equation}
\langle e^{\int d\lambda(\xi_\lambda\hat{A}_\lambda+\eta_\lambda\hat{A}_\lambda^\dagger)}\rangle_c=e^{-\int d\lambda\frac{\xi_\lambda\eta_\lambda}{2}\coth\frac{\beta\hbar\omega_\lambda}{2}}. \label{factor2}
\end{equation}
Then, we can calculate the persistent amplitude by applying (\ref{factor2}) to $\hat{V}^{(+)}(t)=\int d\lambda \alpha_\lambda\hat{A}_\lambda$ with $\alpha_\lambda=\int_0^T dt\sqrt{\frac{\hbar}{2}}kf(t)\frac{\kappa_\lambda\omega_\lambda\sqrt{\omega_\lambda}}{\eta_-(\omega_\lambda)}e^{-i\omega_\lambda t}$ 
\begin{eqnarray}
&&\langle{\rm T}\{e^{-\frac{i}{\hbar}\int_0^T\hat{H}_I(t)}\}\rangle_c \nonumber \\
&=&\langle{\rm N}\{e^{-\frac{i}{\hbar}\int_0^T\hat{V}(t)}\}\rangle_c e^{-\frac{1}{2\hbar^2}\int_0^T dt_1\int_0^T dt_2\langle 0|{\rm T}\{\hat{V}(t_1)\hat{V}(t_2)\}|0\rangle} \nonumber \\
&&\times e^{-\frac{i}{\hbar}\frac{k}{2}\int_0^T dt f(t)^2} \nonumber \\
&=&\langle e^{-\frac{i}{\hbar}\int_0^T\hat{V}^{(-)}(t)}e^{-\frac{i}{\hbar}\int_0^T\hat{V}^{(+)}(t)}\rangle_c \nonumber \\
&&\times  e^{-\frac{1}{2\hbar^2}\int_0^T dt_1\int_0^T dt_2\langle 0|{\rm T}\{\hat{V}(t_1)\hat{V}(t_2)\}|0\rangle}e^{-\frac{i}{\hbar}\frac{k}{2}\int dt f(t)^2} \nonumber \\
&=&e^{-\frac{1}{2\hbar^2}\int d\lambda|\alpha_\lambda|^2(\coth\frac{\beta\hbar\omega_\lambda}{2}-1)} \nonumber \\
&&\times  e^{-\frac{1}{2\hbar^2}\int_0^T dt_1\int_0^T dt_2\langle 0|{\rm T}\{\hat{V}(t_1)\hat{V}(t_2)\}|0\rangle}e^{-\frac{i}{\hbar}\frac{k}{2}\int_0^T dt f(t)^2}. \nonumber \\
&& \label{phase2}
\end{eqnarray}
In the last line, the first exponential factor is equal to 
\begin{equation}
e^{-\frac{4k^2v^2}{\hbar}\int d\lambda\frac{\kappa_\lambda^2}{|\eta_+(\omega_\lambda)|^2\omega_\lambda}\sin^4\frac{\omega_\lambda T}{4}(\coth\frac{\beta\hbar\omega_\lambda}{2}-1)}, \label{amplitude4}
\end{equation}
 which is real describing the decay of the persistent amplitude for the canonical state and does not contribute to the phase shift. The absolute value (\ref{amplitude4}) is an increasing function of $\beta$, and higher temperature requires smaller $v$ to achieve the quasi adiabatic process.
On the other hand, the remaining exponential factors are the same as (\ref{phase1}).
Hence, the persistent amplitude is independent from the inverse temperature $\beta$ in the double limit $v^2\ll\frac{\hbar}{k^2}/\left( d\lambda\frac{\kappa_\lambda^2\omega_\lambda^3}{|\eta_+(\omega_\lambda)|^2}\frac{\sin^4\frac{\omega_\lambda T}{4}}{\omega_\lambda^4}\coth\frac{\beta\hbar\omega_\lambda}{2}\right)$ and $T\gg\int\left( d\lambda\frac{\kappa_\lambda^2\omega_\lambda^3}{|\eta_+(\omega_\lambda)|^2}\frac{\sin^4\frac{\omega_\lambda T}{4}}{\omega_\lambda^4}\coth\frac{\beta\hbar\omega_\lambda}{2}\right)/\left(d\lambda\frac{\kappa_\lambda^2}{|\eta_+(\omega_\lambda)|^2}\right)$ with non negligible $v^2T$, and the  persistent amplitude is well-approximated by $e^{i\Delta \Theta}$ in (\ref{energyshift3}).      
  
Therefore, the phase $\Theta_\mu$ is identical for all the excited states $|E_\mu\rangle$,
\begin{equation}
\Theta_\mu=\frac{k v^2}{2\hbar}\int d\lambda\frac{\kappa_\lambda^2}{|\eta_+(\omega_\lambda)|^2}T+{\cal O}(1), \label{energyshift6}
\end{equation}
 which is our main result.     
\section{Summary}
We have explored the dynamics of the energy eigenstates for nonequilibrium processes.
In particular, we rigorously calculated the persistent amplitude, which measures a sort of distance between the initial and final states. 
In particular, the phase is common for all the excited states, while the absolute value is an increasing function of the inverse temperature.
Then, the quasi adiabatic processes are characterized by the persistent amplitude (\ref{amplitude4}) in a well-defined double limits of small perturbation $v\rightarrow 0$ and long time $T\rightarrow\infty$ with $v^2T$ kept finite.  
\section*{Acknowledgment}
This work was supported by Grants-in-Aid for Young Scientists (B) (No.\ 26800206) from JSPS, Japan. 
\appendix
\section{Proof of the lemma}
Let us show the lemma $\int d\lambda\frac{\kappa_\lambda^2\omega_\lambda^2}{|\eta_+(\omega_\lambda)|^2}=\frac{1}{k}$.
A slightly different calculation is shown in \cite{Monnai2}.
We note that the dispersion functions satisfy $\eta_-(x)-\eta_+(x)=-\pi i\int d\lambda\delta(x-\omega_\lambda)\kappa_\lambda^2\omega_\lambda^3$. 
Then, we can calculate the coefficient as
\begin{eqnarray}
&&\int d\lambda\frac{\kappa_\lambda^2\omega_\lambda^2}{|\eta_+(\omega_\lambda)|^2} \nonumber \\
&=&\int d\lambda\int_0^\infty dx\delta(x-\omega_\lambda)\frac{\kappa_\lambda^2\omega_\lambda^3}{\eta_-(x)\eta_+(x)x} \nonumber \\
&=&\int_0^\infty\frac{1}{-\pi i}(\frac{1}{\eta_+(x)}-\frac{1}{\eta_-(x)})\frac{dx}{x} \nonumber \\
&=&\int_{-\infty}^\infty\frac{1}{-\pi i}\frac{1}{\eta_+(x)x}dx \nonumber \\
&=&\int_C\frac{1}{-\pi i}\frac{1}{\eta_+(z)z}dz-\lim_{r\rightarrow 0}\int_0^\pi\frac{ri e^{i\theta}}{re^{i\theta}\eta_+(re^{i\theta})}d\theta \nonumber \\
&&-\lim_{R\rightarrow\infty}\int_0^\pi\frac{Ri e^{i\theta}}{Re^{i\theta}\eta_+(Re^{i\theta})}d\theta \nonumber \\
&=&\frac{1}{k}. \label{coefficient1}
\end{eqnarray}
Here, the contour $C$ consists of the real axis $[-\infty,-r]$, $[r,\infty]$, and semi circles on the upper-half plane whose centers are $z=0$ with radii $r$ and $R$,  respectively. 
\begin{figure}
\center{
\includegraphics[scale=0.6]{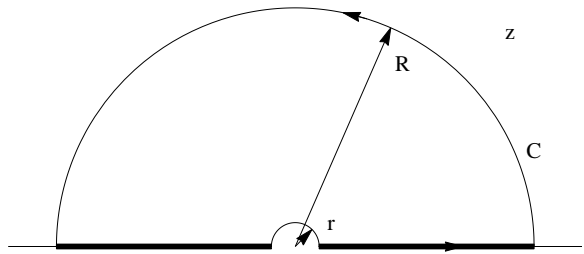}
}
\caption{(Color Online) The contour $C$ on the complex plane consists of $[-\infty,-r]$, $[r,\infty]$, and semi circles on the upper-half plane.}
\end{figure}

\end{document}